\documentclass[twocolumn,showpacs,groupedaddress]{revtex4}
\usepackage{graphicx}

\begin{document}

\title{Dynamical signature of edge state in  the 1D Aubry-Andr\'{e} model}
\author{H. Z. Shen$^{1}$, X. X. Yi$^{1,2}$, and  C. H. Oh$^{2}$}
\affiliation{$^1$School of Physics and Optoelectronic Technology\\
Dalian University of Technology, Dalian 116024 China\\
$^2$Centre for Quantum Technologies and Department of Physics,
National University of Singapore, 117543, Singapore }

\date{\today}

\begin{abstract}
Topological feature has become an intensively studied subject in
many fields of physics. As a witness of topological phase,  the edge
states are topologically protected and may be helpful  in quantum
information processing. In this paper, we define a measure to
quantify the dynamical localization of system and simulate the
localization in the 1D Aubry-Andr\'{e} model. We find an interesting
connection between the edge states and the dynamical localization of
the system, this connection may be used as a signature of edge state
and topological phase.
\end{abstract}

\maketitle
\section{introduction}
The discovery of topological insulators has attracted considerable
attention in the field of topological phases of matter
\cite{Bernevig,Fu,Koenig,Hsieh1,Chen,Xia,Hsieh2,Kuroda}. Earlier
studies   in this field mainly focus on three issues, (1)
topological properties of the ground state phases, (2) realizations
of topological quantum matter\cite{yao13}, and (3) possible
applications of topological matter, e.g., applications of edge
states for topological quantum computing and
spintronics\cite{nayak08,moore10}. Yet, explorations on the
dynamical feature of topological quantum matter are scarce. This
motivates the  study in this paper.

Localization induced by disorder has been recently observed in
ultracold bosonic gases  in purely random potentials \cite{roati08}
and in bichromatic optical lattice\cite{billy08}. In both systems,
the observations have been interpreted in terms of Anderson
localization\cite{anderson58}. Anderson localization is trivial in
one-dimensional systems since the critical disorder at which the
system wavefunction changes from being extended to exponentially
localized is zero, i.e., all states for any finite disorder are
exponentially localized in 1D systems. This makes 1D Anderson
localization rather unattractive. However, by using the so-called
Aubry and Andr\'{e}(A-A)  model,  Aubry and Andr\'{e}\cite{aubry80}
predicted a sharp transition from diffusion to localization for a
given value of disorder length in 1D systems in 80s last century,
where the transition arises from the existence of an incommensurate
potential of finite strength mimicking disorder in a 1D tight
binding model.

The A-A model (also known as Harper model) can be simulated by
trapped fermions on 1D quasi-periodic optical lattice, which can be
generated by superimposing two 1D optical lattice with commensurate
or incommensurate wavelength \cite{fallani07,roati08,deissler10}.
One interesting aspect of the 1D A-A model is that it can be mapped
into the 2D Hofstadter model \cite{harper55,hofstadter76},
exemplifying the topologically-nontrivial 2D quantum Hall system on
a 1D lattice.

In this paper, we focus on the manifestation of topological
properties in the dynamics of topological states in 1D systems. In
the absence of any symmetries, all 1D systems belong to the
topologically trivial  phases, while in 2D systems there are
topological phases of the integer quantum Hall effect
\cite{avron86}. Recently, it has been shown that one-dimensional
quasi-periodic optical lattice systems can exhibit edge states
\cite{lang12} and possess the same physical origins of topological
phases of 2D quantum Hall effects on periodic lattices. This makes
the study of the 1D A-A model rather interesting, which is adopted
as motivation to study the dynamics of the 1D A-A model.

This paper is organized as follows. In Sec. {\rm II}, we introduce
the model and present the equation of motion for the system. In Sec.
{\rm III},  we study the dynamics of the Aubry-Andr\'{e}(A-A) model,
a quantity to characterize the dynamical localization, called
average inverse participation ratio(AIPR), is introduced and
calculated. The dependence of the AIPR on system parameters is given
and discussed. In Sec. {\rm IV}, we study the dynamics of
off-diagonal  Aubry-Andr\'{e}(A-A) model. Finally, we conclude our
results in Sec. {\rm V}.

\section{Diagonal A-A model}

Let us begin with a specific quasi-crystal, the 1D Aubry-Andr\'{e}
(A-A) model~\cite{aubry80}. This is a 1D tight-binding model in
which the on-site potential is modulated in space. The Hamiltonian
of this model takes,
\begin{equation}
H=-J\sum_{i=1}^{N} (\hat{c}^\dagger_i \hat{c}_{i+1}+
\mathrm{H.c.})+\sum_{i=1}^{N} V \mathrm{cos}(2\pi \alpha i+\delta)
\hat{n}_i, \label{H}
\end{equation}
where $N$ is the number of the lattice sites, $\hat{c}^\dagger_i$
($\hat{c}_i$) denotes  the creation (annihilation) operator of the
fermion, and $\hat{n}_i=\hat{c}^\dagger_i \hat{c}_i$.  $J$
represents the hopping amplitude,   $V$ is the modulation amplitude
of the on-site potential, and $\alpha$ controls the periodicity of
the modulation. Whenever $\alpha$ is irrational the modulation is
incommensurate with the lattice and the on-site term is
quasi-periodic. Note that in this model the modulation phase
$\delta$ appears as  an additional degree of freedom   representing
a shift of the origin of the quasi-periodic order. We adopt open
boundary conditions with $n=1$ and $n=N$ being the two edge sites.

Suppose there is only one excitation in the 1D lattice,   the
wavefunction of the system at time $t$ can be written as $|\Psi(t)
\rangle=\sum_{n}^N \psi_{n}(t) c_{n}^{\dagger} |0\rangle$.
Substituting this wavefunction into the Schr\"{o}dinger equation
leads to the following   equation:
\begin{eqnarray}
i\frac{\partial}{\partial
t}\psi_{n}(t)&=&-J(\psi_{n+1}(t)+\psi_{n-1}(t))\nonumber\\
&+&V\cos\left(2\pi \alpha
n+\delta\right)\psi_{n}(t)\,.\label{aamodel}
\end{eqnarray}
Here, $\psi_{n}(t)$ is the probability amplitude of finding the
excitation  at site $n$. For irrational $\alpha$, it is shown that a
localization transition appears in the A-A model as $V$ is increased
beyond the critical value ($V=2J$) with all states being extended
(localized) for $V<2J$ ($V>2J$). For rational $\alpha$, the A-A
model can be mapped into a 2D Hofstadter lattice by treating
$\delta$ as the momentum of another spatial
dimension\cite{harper55,hofstadter76}. For $\alpha\neq \frac 1 2 $,
the Hofstadter lattice has gapped energy bands with non-trivial
topology, characterized by non-zero Chern numbers. Thus localized
edge states are expected for a finite-size system with boundary.
\begin{figure}
\includegraphics*[width=0.8\columnwidth,
height=0.6\columnwidth]{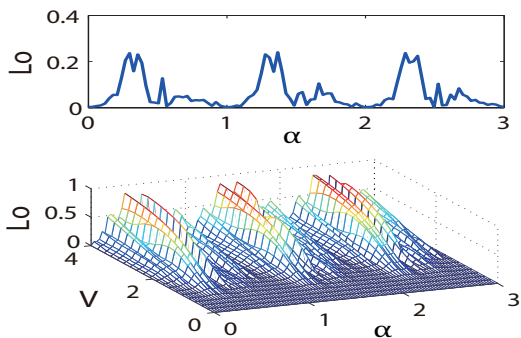} \caption{The average inverse
participation ratio as a function of $\alpha$ with a given $V=1.5J$
(top panel), and as a function of $\alpha$ and $V$ (bottom panel).
The total number of sites is $N=99$. The other parameters chosen are
$J=1$,$\delta=\frac{2\pi}{3}$. The system is initially at site 1.
$V$ is chosen in units of $J$.} \label{fig1}
\end{figure}
\section{results}
To simulate numerically the quantum evolution of the system, we
choose two different initial states. In the first, the system is
initially prepared to occupy site 1, while in the second, the system
is initialized at the center of the lattice. To quantify the
dynamical localization/extension of the  system, we define an
average inverse participation ratio(AIPR),
\begin{equation}
Lo=\frac 1 T\int_0^Tdt \sum_n|p_n(t)|^2,
\end{equation}
where $p_n(t)$ denotes the probability of finding the excitation at
site $n$, therefore, $\sum_np_n(t)=1$. $T$ denotes the evolution
time. This definition can be understood as an extension of the
inverse participation ratio averaged over the evolution time $T$.
Therefore, the AIPR depends on the initial state of the system. In
the following numerical simulation, we initialize the system to
occupy one of the edge sites, say site 1, at the beginning of
evolution. Due to the exchange symmetry of the 1D system, the edge
eigenstates must appear in pairs or occupy the two edge sites with
equal probability. In other words, when we find an edge state
located at edge site 1, there must be another one at the edge site
N. Otherwise, the edge eigenstate distributes equally at both edge
sites. Assume the edge state is $c_1^{\dagger}|0\rangle$, i.e., the
edge eigenstate is exactly the excited state of site 1, it is easy
to show that $Lo=1.$ For edge states that not only locate exactly at
the edge sites, the probability of finding the system at site 1
after an evolution time $t$ is $P_1(t)=|\langle
1|\Phi(t)\rangle|^2$, where $|\Phi(t)\rangle$ is the wavefunction of
the system at time $t$ with initial state $|1\rangle$.
Straightforward calculation yields,
$$P_1(t)=\sum_n |a_n|^4+2\sum_{m\neq n} |a_n|^2|a_m|^2
\cos(E_m-E_n)t,$$ where $a_n$ denotes a coefficient defined by
$|1\rangle=\sum_na_n|E_n\rangle$, and $|E_n\rangle$ is the n-th
eigenstate of the system with eigenvalue  denoted by $E_n$.
Therefore,
\begin{equation}
Lo=\frac 1 T\int _0^T P_1^2(t) dt\sim |a_1|^8+4|a_1|^4\sum_{m,n\neq
1}|a_n|^2|a_m|^2,
\end{equation}
where $|E_1\rangle$ being the edge state located at the edge site 1
is assumed. Thus, for a system having edge eigenstates,
$Lo\rightarrow 1,$ whereas $Lo\rightarrow 0,$ for system that all
eigenstates are extended states. Thus the AIPR can be taken as a
measure to quantify the localization of the dynamics.

Fig.\ref{fig1} displays the average inverse participation ratio $Lo$
as a function of $V$ and $\alpha$ (lower panel). To show clearly the
dependence of $Lo$ on $\alpha$, we present  the $Lo$ versus $\alpha$
with a fixed $V=1.5$ in the upper panel. We find that $Lo$ arrives
at its maximum at about ($\alpha=$integer+$\frac 1 3$). As $V$ is
increased, the system becomes more dynamically localized at the edge
sites. $Lo$ is a periodic function of $\alpha$ and $\delta$ with
periods 1 and $2\pi$, respectively. This is a reflection of symmetry
in the Hamiltonian, i.e., the Hamiltonian remains unchanged by
substitution, $\alpha\rightarrow (1+\alpha)$, $\delta\rightarrow
(\delta+2\pi).$ From Fig. \ref{fig1} we can also observe that $Lo$
is very close to zero at $\alpha=m$ and $\alpha=\frac {(2m+1)} {2}$
where $m$ is an integer. This can be explained as a direct
consequence of the space-independent on-site potential,
$V_i=V\cos(2\pi\alpha i+\delta)|_{\alpha=m}=V\cos\delta$ and
$V_i|_{\alpha=\frac {(2m+1)}{2}}=-\cos\delta.$

\begin{figure}
\includegraphics*[width=0.8\columnwidth,
height=0.8\columnwidth]{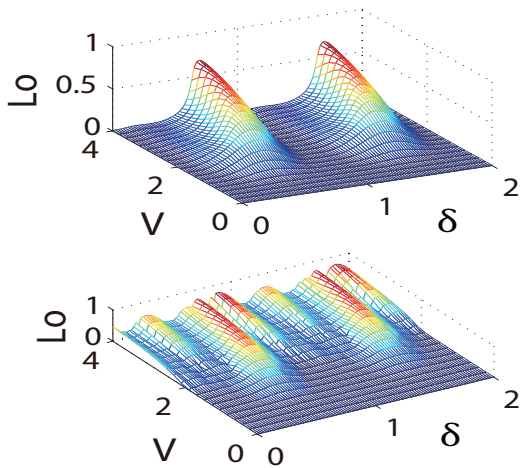} \caption{The AIPR as a
function of $V$ and $\delta$. The total number of sites is $N=99$,
the other parameters chosen are $J=1$. For the top panel,
$\alpha=\frac 1 3$, and for the bottom panel $\alpha=(\sqrt 5-1)/4$.
The system is initially prepared at site 1. $V$ is in units of $J$,
$\delta$ is in units of $\pi$.} \label{fig2}
\end{figure}

\begin{figure}
\includegraphics*[width=0.8\columnwidth,
height=0.5\columnwidth]{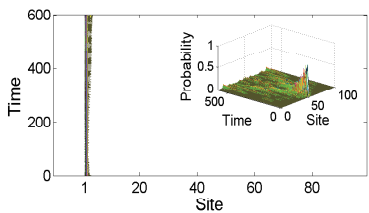} \caption{Probability
distribution over sites at each instance of time. The system is
initially at site 1. The total number of site is $N=99$. $J=1,
V=2.2J, \alpha=0.3, \delta=\frac{2\pi}{3}$. The evolution time is
$T_{max}=600$ (in units of $\frac 1 J$). The inset is for the system
initially at the site 50, i.e., the center of the lattice. This
figure shows that the system is well localized when it is initially
in site 1, but it is not dynamically localized when the system is
initially at site 50.} \label{fig3}
\end{figure}

Fig.\ref{fig2} shows the AIPR as a function of $V$ and $\delta$. As
$\alpha$ changes, $Lo$ may have one peak or many peaks within one
period of $\delta$, each peak corresponds to an eigenstate well
localized at the edge site 1.
The observed dynamical localization quantified by AIPR depends on
the initial state of the system, for example, the system is well
dynamically localized with the excitation being initially at site 1,
while it is extended with the central site being occupied, see Fig.
\ref{fig3}. Further numerical results show that $Lo$ is sharply
suppressed when the sites other than site 1 are occupied initially.

\begin{figure}
\includegraphics*[width=0.8\columnwidth,
height=0.6\columnwidth]{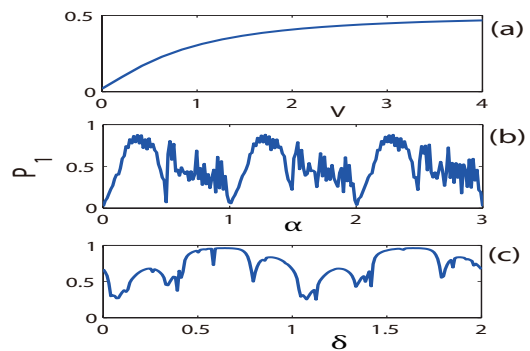} \caption{The largest
probability of finding the system at site 1, when the system  in one
of its eigenstates. $N=99$ and $J=1$. (a)The probability versus $V$
with  $\alpha=1/3$ and $\delta=\frac{2\pi}{3}.$ (b)The probability
as a function of $\alpha$ with $V=2.2,$ and $\delta=\frac{2\pi}{3}.$
(c)As a function of $\delta$ with $\alpha=(\sqrt{5}-1)/4,V=4.$}
\label{fig4}
\end{figure}

To understand the physics behind the dynamical localization, we
calculate the largest probability of finding the system at site 1
when the system is in one of its eigenstates. This calculation would
show the overlap between the site 1 and the eigenstate which
exhibits largest probability to occupy site 1. The results are given
in Fig. \ref{fig4}. Comparing Fig. \ref{fig1}, Fig. \ref{fig2} and
Fig. \ref{fig4}, we find that the $Lo$ and $P_1$ reach their
respective maximum and minimum at almost the same $\alpha$ and
$\delta$. The dependence of $Lo$ and $P_1$ on $V$ manifests the same
feature, i.e., they increase as $V$ increases.

To shed light on the effect of boundary, we now turn to discuss the
situation with a periodic boundary condition. The A-A model in this
situation can be solved analytically as follows. Suppose that the
n-th eigenstate of the Hamiltonian restricted to a single particle
in the 1D lattice is given by $|E_n\rangle =\sum_i u_{i,n}c_i^\dag|
0\rangle$, the eigenvalue equation $H| E_n\rangle = E_n|E_n\rangle$
leads to the Harper equation,
\begin{eqnarray}
E_nu_{i,n}&=&-J(u_{i + 1,n} + u_{i - 1,n})\nonumber\\
&+&V\cos(2\pi\alpha i + \delta )u_{i,n}. \label{eigenstate}
\end{eqnarray}
We shall consider the commensurate potential $V_i=V\cos(2\pi\alpha
i+\delta)$ with a rational $\alpha $ given by $\alpha  = p/q$ with
$p$ and $q$ being integers which are prime to each other. Since the
potential $V_i$ is periodic with a period $q$, the wave functions
take the Bloch form,  ${u_{i + q}} = {e^{ikq}}{u_i}$, for the
lattice with the periodic boundary condition. Omitting the
eigenstate index $n$ where not confused  and taking $u_j =
e^{ikj}{\varphi_j}(k)$ for $|k| \le \pi /q$, we have $\varphi_{j +
q}(k)=\varphi_j(k)$, Eq.~(\ref{eigenstate}) then follows
\begin{eqnarray}
E(k)\varphi_j&=&-J(e^{ik}\varphi_{j + 1} + e^{ - ik} \varphi _{j -
1})\nonumber\\
&+&V\cos(2\pi jp/q +\delta)\varphi_j. \label{momentum}
\end{eqnarray}
To be specific, we take $\alpha = 1/3$, then $q=3$ and $- 1/3 \le
k/\pi \le 1/3$. By $\varphi_{j + q}(k)=\varphi_j(k)$,
Eq.~(\ref{momentum}) reduces to,
\begin{eqnarray}
M\Phi_E = E\Phi_E, \label{fr}
\end{eqnarray}
where
\begin{eqnarray}
M = \left(
\begin{array}{ccc}
V\cos (\delta+\frac{2\pi}{3}) & -J{e^{ik}}  &  - J{e^{ - ik}}\\
-Je^{ - ik}  &  V\cos (\delta-\frac{2\pi}{3} ) & - J{e^{ik}}\\
-J{e^{ik}}  & -Je^{-ik} & V\cos \delta
\end{array}
\right),\nonumber\\
\end{eqnarray}
and $\Phi_E= \left( \varphi_1, \varphi_2, \varphi_3\right)^T$.
Solving the eigenvalue equation (\ref{fr}), we can obtain three
eigenvalues for a given $k$,
\begin{eqnarray}
{E_1} &=& \sqrt{4J^2+V^2}\cos \theta,\nonumber\\
{E_2} &=& \sqrt{4J^2+V^2}\cos (\theta+ \frac{{2\pi }}{3}),\nonumber\\
{E_2} &=& \sqrt{4J^2+V^2}\cos (\theta- \frac{{2\pi }}{3}),
\end{eqnarray}
where $\theta=\frac 1 3 \arccos (-\frac{d}{2\sqrt{(J^2+\frac 1
4V^2)^3}}),$ $d = 2J^3\cos (3k ) - \frac{V^3}{4}\cos (3\delta).$ The
corresponding three eigenstates are,
\begin{eqnarray}
\Phi_E= \left(\begin{array}{r}
\varphi_1(E)\\
\varphi_2(E)\\
\varphi_3(E)
\end{array}\right),\ \ E=E_1,E_2,E_3,
\end{eqnarray}
where
\begin{eqnarray}
\varphi_1(E)&=& g(E)\varphi_2,\nonumber\\
\varphi_2(E)&=& \frac{1}{\sqrt {1 + |g(E)|^2 + |e(E)|^2}},\nonumber\\
\varphi_3(E)&=& {e(E)}{\varphi_2},\nonumber\\
g(E)&=& \frac{e^{ik } + B(E)e^{ - 2ik }}{A(E)-B(E)},\nonumber\\
e(E)&=& B(E)\cdot\frac{e^{2ik }+A(E)e^{-ik}}{A(E)- B(E)},\nonumber\\
A(E)&=&\frac{V\cos(\frac{2\pi}{3}+\delta)-E}{J},\nonumber\\
B(E)&=&\frac{J}{V\cos\delta-E}.
\end{eqnarray}
Collecting these equations, we calculate the AIPR and exhibit  the
results in Fig. \ref{fig5} (bottom panel). The AIPR under the
periodic condition is very small,  by contrast, the open boundary
condition helps dynamical localization, as the top panel of Fig.
\ref{fig5} shows. This again can be explained as a consequence of
small overlap between the site 1 and the eigenstates of the system.
In fact, under the periodic condition, all eigenstates of the system
are extended, this is enforced by the Bloch theorem.
\begin{figure}
\includegraphics*[width=0.8\columnwidth,
height=0.8\columnwidth]{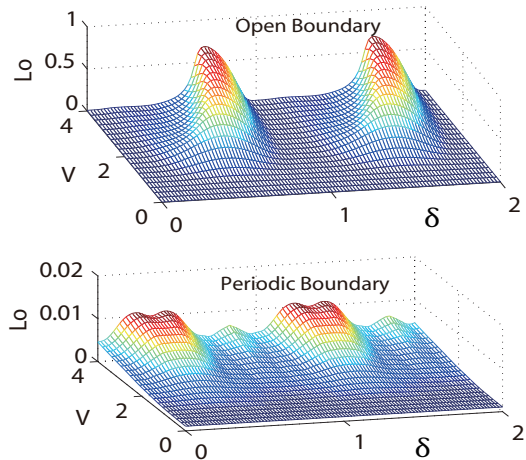} \caption{The AIPR as a
function of $V$ and $\delta$ for different boundary conditions. Top
panel is for open boundary condition, while the bottom is for
periodic boundary condition. The system is initialized  at the edge
site 1. $N=99$,  $J=1$, and $\alpha=1/3.$} \label{fig5}
\end{figure}

The time evolution of an isolated macroscopic quantum system
initially prepared in an out-of-equilibrium state is currently
turning from an abstract concept to a real phenomenon that can be
observed and studied experimentally.  This striking change has been
mainly driven by experiments on cold
atoms\cite{bloch08,polkovnikov11},  but it will be surely given
further impulse in the near future by the fast progresses in
time-resolved spectroscopy on condensed-matter systems.

\section{off-diagonal A-A model}
Now we extend the study to the the generalized 1D A-A model, which
is described by the following Hamiltonian,
\begin{eqnarray}
H&=&-J\sum_{i=1}^{N}\left [ 1+\lambda\cos(2\pi \alpha
i+\delta_{off})\right ] (\hat{c}^\dagger_i \hat{c}_{i+1}+
\mathrm{H.c.})\nonumber\\
&+&\sum_{i=1}^{N} V \mathrm{cos}(2\pi \alpha i+\delta) \hat{n}_i.
\label{Hoff}
\end{eqnarray}
This Hamiltonian is different from the model  in Eq. (\ref{H}) at
the inhomogeneity in hopping strength described by cosine
modulations. The  modulations have the same periodicity as in the
on-site potential energy and its amplitude is characterized by
$\lambda$. The special case with $\lambda=0$ corresponds to the
diagonal A-A model, and the generalized A-A model can be derived
starting from an ancestor 2D Hofstadter model with
next-nearest-neighbor hopping terms. It has been shown recently
\cite{sarma13} that the commensurate off-diagonal A-A model is
topologically nontrivial in the gapless regime and supports
zero-energy edge modes. Unlike the incommensurate case, the
nontrivial topology in the off-diagonal A-A model is attributed to
the topological properties of the one-dimensional Majorana chain.

\begin{figure}
\includegraphics*[width=0.8\columnwidth,
height=0.8\columnwidth]{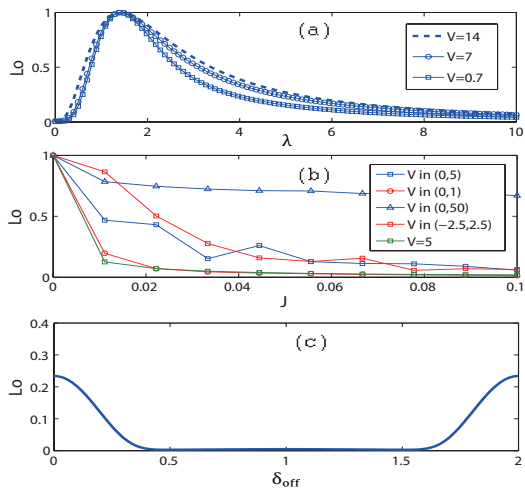} \caption{The AIPR as a
function of $\lambda$, $J$ and $\delta$ for $N=100$ sites. (a)$J=1$,
$V=14, 7, 0.7$(from dashed to square),  $\alpha=1/2,$
$\delta=2\pi/3,$ $\delta_{off}=\pi/4.$ (b) $\lambda=0,$
$\alpha=1/2,$ $\delta=2/3\pi,$ $\delta_{off}=\pi/4,$ (c) $J=1,$
$\alpha=1/2,$ $\delta=0,$ $\lambda=0.4,$ $V=0.$} \label{fig5}
\end{figure}

Fig. \ref{fig5} shows the AIPR as a function of $\lambda$,  $J$ and
$\delta_{off}$ with the system initially at site 1. We find that
$V$, the modulation amplitude of the on-site potential, does not
shift the peak, but the larger the $V$, the bigger  the AIPR is
(Fig. \ref{fig5}-(a)). Randomness of  $V$ increases the AIPR, see
Fig. \ref{fig5}-(b), which is  reminisant of Anderson localization
in a disordered medium. The phases $\delta$ and $\delta_{off}$ can
be tuned independently in experiment, so  $\delta$ and
$\delta_{off}$ can be treated as independent variables. The phase
$\delta_{off}$ can alter the AIPR, see Fig. \ref{fig5}-(c), the AIPR
 arrives at a maximum when $\delta_{off}=0$, and it is very close to
0 when $\delta_{off}$ in an interval of  $[\pi/2, 3\pi/2]$.

\section{Conclusion}
To conclude, we have found a signature for the existence of edge
states in terms of dynamical feature of the system. A quantity to
measure the dynamical localization called average inverse
participation ratio (AIPR) is introduced and discussed. We have
calculated the AIPR for the diagonal and off-diagonal
diaAubry-Andr\'{e} model. Our findings suggest that the AIPR can be
taken as a measure to quantify the dynamical localization of quantum
system, in particular it can be chosen as a witness of edge state.
Our strategy is to make use of the definition of edge states and it
is applicable for both gapped and gapless systems in one dimension.

\ \ \\
This work is supported by the NSF of China under Grants Nos
61078011,  10935010 and 11175032 as well as the National Research
Foundation and Ministry of Education, Singapore under academic
research grant No. WBS: R-710-000-008-271.

\end{document}